\documentclass[aps,prd,twocolumn,groupedaddress,nofootinbib]{revtex4-2}
\usepackage{tensor, amsmath, amssymb, graphics, geometry, physics, mathtools, graphicx, wrapfig}
\usepackage[
	colorlinks=true,
	linkcolor=blue,
	citecolor=blue,
	filecolor=magenta,
	urlcolor=blue
]{hyperref}
\newcommand{\barT}{{\overline{T}}}

\begin{document}

\title{Comparison of \textit{f}(\textit{R},\textit{T}) Gravity with Type Ia Supernovae Data}
\author{Vincent R. Siggia}
\author{Eric D. Carlson}
\email{ecarlson@wfu.edu}
\affiliation{Department of Physics, Wake Forest University, 1834 Wake Forest Road, Winston-Salem, North Carolina 27109, USA}

\begin{abstract}
    The expansion of the Universe in $f(R,T)$ gravity is studied. By focusing on functions of the form $f(R,T)=f_1(R)+f_2(T)$, we assert that present-day acceleration can be achieved if the functional form of $f_2(T)$ either grows slowly or falls as a function of $T$. In particular, we demonstrate that when $f_2(T) \propto T^{-1}$, the Universe transitions to exponential growth at late times, just as it does in the standard cosmological model. A comparison of predictions of this model with type Ia supernovae shows that this model fits the data as well or even slightly better than the standard cosmological model without increasing the number of parameters.
\end{abstract}

\maketitle

\section{Introduction}

For over two decades, we have known that the expansion of the Universe is accelerating. Originally this was shown by studying type Ia supernovae (SNe Ia) \cite{Riess_1998,Perlmutter_1998,Perlmutter_1999}. Measurements of the cosmic microwave background radiation \cite{Planck2018} and baryon acoustic oscillations show the same thing \cite{Alam_2017}.

The simplest and most common explanation is to assume that there is a cosmological constant $\Lambda$, such that the gravitational action normally proportional to the curvature scalar $R$ is modified to $R +2\Lambda$. The evidence \cite{Planck2018} indicates that the Universe is close to spatially flat, and the resulting $\Lambda$ cold dark matter ($\Lambda$CDM) model fits well with available data. 

However, many alternatives have been considered, including modifications of gravity. In $f(R)$ gravity, the curvature term $R$ is modified to be some function of the curvature scalar \cite{10.1093/mnras/150.1.1}. Another alternative, proposed first by Harko \textit{et al.}, is to consider a contribution of the form $f(R,T)$, where $T$ is the trace of the stress-energy tensor \cite{Harko_2011}. 

To find $T$, it is first necessary to derive the stress-energy tensor from the matter Lagrangian $\mathcal{L}_m$. When dealing with ordinary matter or dark matter, the matter Lagrangian comes from either the standard model Lagrangian or whatever extension is responsible for dark matter, thus making the explicit form of the stress-energy tensor either complicated or unknown. A common strategy in both standard general relativity and in these modified theories is to assume a perfect fluid stress-energy tensor. The method for doing this in general relativity is well known \cite{Brown_1993}, but there are additional complications that occur in modified gravity, as pointed out by \cite{Minazzoli_2012} and later elaborated by \cite{Fisher_2019}. Multiple papers \cite{Harko_2011,S_ez_G_mez_2016,PhysRevD.94.084052,PhysRevD.95.123536,Carvalho_2017,Deb_2018} ignore these complications, making their conclusions suspect. One attempt to bypass the ambiguity is proposed by \cite{Carvalho_2021}. However, this attempt contained some sign inconsistencies, partially due to errors in the literature related to the metric conventions, and a detailed analysis of their work leads us to conclude that this approach is not productive.

One of the simplest examples that can be considered is when $f(R,T)$ is additively separable, such that
\begin{equation}\label{additively separable}
    f(R,T) = f_1(R) + f_2(T)\;.
\end{equation}
As pointed out by \cite{Fisher_2019}, in such theories, $f_2(T)$ can, in principle, be incorporated into the matter Lagrangian $\mathcal{L}_m$ and, as such, perhaps should not be considered as a modification of gravity at all. This is indeed the argument of \cite{Lacombe_2024}. However, whether this is considered as a modification of gravity or not, we can still ask the question of whether such a theory can account for the accelerating expansion of the Universe.

In this paper, we focus on the question of how one can reproduce a currently accelerating Universe with a theory of the form of Eq.~(\ref{additively separable}). Since we are focusing on the effects of the stress-energy term, we use $f_1(R) = R$ and use the simplest possible form $f_2(T) = \lambda T^\epsilon$. As pointed out by \cite{Lacombe_2024}, if we choose $\epsilon = 2$, then standard model constraints at the weak scale places strong limits on $\lambda$. At the much lower energy densities relevant in current-day cosmology, the new term would be irrelevant compared to other terms in the action such as $\mathcal{L}_m$. This argument generalizes to conclude that any $\epsilon > 1$ is irrelevant for present-day cosmology, and we will focus on $\epsilon < 1$. The case when $\epsilon=0$ corresponds to the standard $\Lambda$CDM. For comparison we will choose $\epsilon=-1$, which should have an effect on the late-time Universe. We also assume the Universe is flat, so as to not introduce unnecessary additional parameters. Thus our alternative has the same number of parameters as standard $\Lambda$CDM, since the cosmological constant $\Lambda$ is replaced by the new coupling constant $\lambda$.

In Sec.~\ref{sec:f(R,T)}, the $f(R,T)$ formalism is laid out, yielding the modified Einstein's equations. In Sec.~\ref{sec:Fluids}, perfect fluids as discussed by \cite{Fisher_2019,Brown_1993} are reviewed for the particular case of $f(R,T)=R+\lambda T^\epsilon$. Then in Sec.~\ref{sec:Scale}, the scale factor of the Universe as a function of time is derived for our model and compared to $\Lambda$CDM results. Finally in Sec.~\ref{sec:Data}, SNe Ia data from the Pantheon dataset \cite{Scolnic_2018,Lu_2022,Abbott_2019} are compared against  both our model and $\Lambda$CDM.

Throughout, we use conventions where $c=\hbar=1$, the signature of the metric is $(+,-,-,-)$, and the Riemann and Ricci Tensors are defined by ${R^\alpha}_{\mu\beta\nu} = \nabla_\beta \Gamma^\alpha_{\mu\nu} - \cdots$ and $R_{\mu\nu}={R^\alpha}_{\mu\alpha\nu}$ respectively.

\section{f(R,T) Formalism}\label{sec:f(R,T)}

In $f(R,T)$ gravity, the Ricci scalar $R$ appearing in the Einstein-Hilbert action is replaced by an arbitrary function $f(R,T)$, where $T$ is the trace of the stress-energy tensor, to yield an action given by
\begin{subequations}
\begin{align}
    S&=\int d^4 x \sqrt{-g}\mathcal{L} \; ,\\
    \mathcal{L}&=\mathcal{L}_m-\frac{1}{2\kappa^2}f(R,T)\;, \label{FullLagrangian}
\end{align}
\end{subequations}
where $\kappa^2=8\pi G$. Varying the action with respect to the metric and identifying the stress-energy tensor as
\begin{equation}
    T_{\mu\nu}=\frac{2}{\sqrt{-g}}\frac{\delta S_m}{\delta g^{\mu\nu}}
\end{equation}
leads to the $f(R,T)$-Einstein's equations
\begin{equation}\label{Einstein EQ}
    \left(R_{\mu \nu }+g_{\mu \nu }\,\square -\nabla _{\mu }\nabla _{\nu }\right)f_R-\frac{1}{2}f\, g_{\mu \nu }=\kappa ^2T_{\mu \nu }-f_T\frac{\partial\,T}{\partial g^{\mu \nu }}\;,
\end{equation}
where the subscripts on $f$ denote partial derivatives \cite{Harko_2011}. Standard gravity with a cosmological constant $\Lambda$ can be recovered by choosing $f(R,T)=R+2\Lambda$. Note that the stress-energy is not conserved in these models \cite{Harko_2011,PhysRevD.90.028501}\footnote{Note a missing term from \cite{Harko_2011} is corrected in \cite{PhysRevD.90.028501}}. Taking the divergence of Eq.~(\ref{Einstein EQ}) gives
\begin{equation}
    \kappa^2\nabla^\mu T_{\mu\nu}= \nabla^\mu\left(f_T\frac{\partial\,T}{\partial g^{\mu \nu }}\right)-\frac{1}{2}f_T\nabla_\nu T\;.
\end{equation}

\section{Perfect Fluids}\label{sec:Fluids}
In general relativity, one is rarely interested in the detailed fundamental Lagrangian. We would prefer to treat the matter content as a perfect fluid. A perfect fluid is described in terms of its local four-velocity $u^\mu$ normalized so that $u^\mu u_\mu=1$, the comoving number density $n$, and the entropy per particle $s$. We expect both the particle number and entropy must be conserved, i.e. 
\begin{subequations}
    \begin{align}
        0&=\nabla_\mu(n\,u^\mu)\;, \label{ConserveNumber}\\
        0&=\nabla_\mu(s n\,u^\mu)\;. \label{ConserveEntropy}
    \end{align}
\end{subequations}

The stress-energy tensor is given by
\begin{equation}
    T_{\text{PF}}^{\mu\nu}=(\rho+p)u^\mu u^\nu-p g^{\mu\nu}\;,
\end{equation}
where $\rho=\rho(n,s)$ is the energy density and $p=p(n,s)$ is the pressure. If stress-energy is conserved then, using the equation $\nabla_\mu T^{\mu\nu}=0$, it can be shown that the energy density and pressure are related by
\begin{equation}\label{pressure relation}
    n\frac{\partial\rho }{\partial n}=\rho+p.
\end{equation}
We will discover when considering nonstandard gravity that the naïve number density and stress-energy tensor are not always conserved; nonetheless, we will treat Eq.~(\ref{pressure relation}) as a definition for the naïve pressure $p$.

In standard gravity, it is common to assume $\mathcal{L}_m=p$ \cite{Harko_2011}, but it is worth understanding the origin of this expression, which for nonstandard gravity turns out not to be so simple, as was pointed out by \cite{Brown_1993}. We start with the form
\begin{equation}\label{General Lagrangian}
    \mathcal{L}_m=-\rho (n,s)+J^{\mu }\left(\beta _A\nabla _{\mu }\alpha ^A-s\nabla _{\mu }\theta -\nabla _{\mu }\phi \right)\;,
\end{equation}
where $J^\mu=n u^\mu$ is the current density, $\alpha^A$ are a set of index functions used to label fluid flow line, and $\beta_A,\theta$ and $\phi$ are Lagrange multipliers used respectively to ensure that current flows along the flow lines, entropy is conserved, and the current is conserved. The number density $n$ is now to be interpreted as an implicit function of $J^\mu$, given by
\begin{equation}\label{Particle Number}
    n=\sqrt{g_{\mu \nu }J^{\mu }J^{\nu }}.
\end{equation}

Working with the full Lagrangian Eq.~(\ref{FullLagrangian}), the stress-energy tensor and its trace are given by
\begin{subequations}
    \begin{align}
         \label{General Stress-Energy}T_{\mu \nu }&=(\rho +p)u_{\mu }u_{\nu }-g_{\mu \nu }\mathcal{L}_m \; , \\
        \label{General Stress-Energy Trace}T&=(\rho +p)-4\mathcal{L}_m\;,
    \end{align}
\end{subequations}
while the variation of ${T}$ with respect to the metric can be determined, using Eq.~\ (\ref{Particle Number}), as
\begin{equation}\label{T Variation}
    \frac{\partial T}{\partial g^{\mu\nu }}=-\frac{1}{2}u_{\mu }u_{\nu }\left(4+n\frac{\partial }{\partial n}\right)(\rho +p)\;.
\end{equation}

Considering the additively separable form Eq.~(\ref{additively separable}) and the equations of motion of Eq.~(\ref{General Lagrangian}) \cite{Brown_1993,Fisher_2019}\footnote{Note that a sign error from \cite{Fisher_2019} has been corrected in Eq.~(\ref{sign error})}
\begin{subequations}
\begin{eqnarray}
0 &=& \nabla\!_\mu \left\{ \left[1 +2\kappa^{-2} f_2^\prime(T)\right] J^\mu \right\}\; , \label{ConserveN} \\
 0 &=& \nabla\!_\mu\left\{s\left[1 + 2\kappa^{-2} f_2^\prime(T)\right]J^\mu\right\} \; , \label{ConserveS} \\
0 &=& \left[1 + 2\kappa^{-2} f_2^\prime(T)\right]J^\mu \nabla\!_\mu\alpha^A \; ,\\
0 & =&  -\nabla\!_\mu\left\{\beta_{\!A}\left[1 + 2\kappa^{-2} f_2^\prime(T)\right]J^\mu\right\} \; ,\\
0 &=& -\left[1 + \frac{2}{\kappa^2} f_2^\prime(T)\right]\left[\frac{\partial\rho}{\partial s}+J^{\mu}\nabla\!_\mu\theta\right] \nonumber \\
&&\qquad {} - \frac{1}{2 \kappa^2} f^\prime_2(T) \frac{\partial}{\partial s}(\rho+p) \label{sign error}\; ,\\
0 &=&\left[1 + \frac{2}{\kappa^2} f_2^\prime(T)\right] \left( \beta_A \nabla_\mu \alpha^A - s\nabla_\mu \theta - \nabla_\mu \phi  - \frac{\partial \rho}{\partial n} u_\mu \right)  \nonumber \\
&&\qquad {} - \frac{1}{2\kappa^2}  f^\prime_2(T) u_\mu \frac{\partial}{\partial n} \left(\rho+p\right) \; . \label{StationarityJ}
\end{eqnarray}
\end{subequations}
We immediately note from Eqs.~(\ref{ConserveN}) and ~(\ref{ConserveS}) that the naïve number density $n$ and entropy density $s$ are \textit{not} conserved. One can obtain the ``on-shell'' matter Lagrangian to be
\begin{equation}\label{lm-bar}
    \overline{\mathcal{L}_m}=p+\frac{\overline{f_2'}}{2\kappa ^2+4\overline{f_2'}}n\frac{\partial }{\partial n}(\rho +p)\;,
\end{equation}
where the bars mean that the functions are to be evaluated using Eq.~(\ref{StationarityJ}) to eliminate all variables except the number density $n$ and entropy per particle $s$. It is then trivial to see that in standard gravity, $\mathcal{L}_m = p$ will occur, but this equation is not valid when $f_2(T)$ is nontrivial.

In order to reproduce Einstein's equations, the function must be $f(R,T)=R$. The next simplest version of $f(R,T)$ should also contain contributions from the stress-energy tensor, in the form of its trace. A suitable additively separable function should be of the form
\begin{equation}\label{f(R,T)}
    f(R,T)=R+\lambda T^\epsilon\;,
\end{equation}
where $\lambda$ is a coupling constant and $\epsilon$ is an arbitrary real number. The choice of the exponent $\epsilon$ will govern how the modified theory will differ from conventional gravity. In ordinary gravity, $T \propto R$, and therefore if $\epsilon > 1$, we should not be surprised to find that the modification is most important at early times when the density is high. What we want is a contribution that will cause significant changes only in the late Universe when densities are low, which can account for the currently accelerating Universe, which suggests we should try $\epsilon < 1$. In particular, for $\epsilon = 0$, the ``new'' term corresponds to simply a cosmological constant with $\lambda = 2\Lambda$. To explore a truly new scenario, we instead focus on other values of $\epsilon$.

Continuing with this form of function, combining Eqs.~(\ref{General Stress-Energy}) and (\ref{lm-bar}), the ``on-shell" stress-energy tensor becomes
\begin{align}
         \label{Stress-Energy}\barT_{\mu \nu }&=T_{\mu \nu }^{\text{PF}}-g_{\mu \nu }\frac{\epsilon\lambda }{2\kappa ^2\barT^{1-\epsilon}+4\epsilon\lambda }n\frac{\partial }{\partial n}(\rho +p)\;.
\end{align}
Taking the trace, using Eq.~(\ref{pressure relation}), and rearranging yields
\begin{equation}\label{General Quad}
    0=T^{\text{PF}}-\barT- \frac{2\lambda\epsilon }{\kappa ^2}\barT^{\epsilon }-\frac{2\lambda\epsilon }{\kappa ^2}\barT^{\epsilon -1}\left(4+n\frac{\partial }{\partial n}\right)p\;.
\end{equation}
This equation should be thought of as an implicit equation for $\barT$ in terms of $n$ and $s$.

\section{The Scale Factor}\label{sec:Scale}

The initial motivation for considering dark energy was the study of SNe Ia. A plot of the luminosity distance versus redshift $z$ seemed to indicate that, unlike a matter- or curvature-dominated universe, the Universe was currently accelerating its expansion. It is useful to understand how this comes about in the standard $\Lambda$CDM model, with $f(R,T) = R+2\Lambda$. We assume a spatially flat Universe, with metric
\begin{equation}\label{FLRW Metric}
    ds^2=d t^2-a^2(t)\left(d r^2+r^2d\Omega^2\right)\;,
\end{equation}
where $a$ is an arbitrarily normalized scale factor. It is evident from Eq.~(\ref{ConserveN}) that in this case the number density $n$ is conserved, and in the nonrelativistic current Universe we would have $\rho \propto n$, so that $\rho a^3$ will be a constant. The $tt$-component of Eq.~(\ref{Einstein EQ}) is then
\begin{equation}
    3\left(\frac{\dot a}{a}\right)^2=\kappa^2\rho+\Lambda \;.
\end{equation}
This equation can be solved exactly to yield $a \propto \sinh^{2/3}(\frac{3}{2} H_\Lambda t)$, where $H_\Lambda=\sqrt{\frac13\Lambda}$, which up to an arbitrary normalization constant has the asymptotic forms
\begin{equation}\label{Asymtotic lCDM}
    a(t)=
    \begin{cases} 
      N(3t)^{2/3} & t\rightarrow 0\;, \\
      NH_\Lambda^{-2/3}\exp(H_\Lambda t)& t\rightarrow\infty\;.
    \end{cases}
\end{equation}

Can we get a similar outcome from $f(R,T)$ gravity? We will assume again that the space metric is flat, given by Eq.~(\ref{FLRW Metric}). The situation is complicated because quantities like the energy and pressure derived \textit{just} from the matter Lagrangian $\mathcal{L}_m$ will not be conserved, nor will be the number density. However, since the energy density $\rho$ and pressure $p$ are being derived simply from the matter Lagrangian, we expect that in the nonrelativistic era, the pressure will still be $p=0$. We expect the naïve energy density $\rho\propto n$ because they are both derived from the standard matter Lagrangian $\mathcal{L}_m$.

To explore this model explicitly, we will focus on the model given by Eq.~(\ref{f(R,T)}), with $\epsilon = -1$. This reduces Eq.~(\ref{General Quad}) to
\begin{equation}\label{Quadratic}
    \rho=\barT- \frac{2\lambda }{\kappa ^2\barT}\;,
\end{equation}
which can be solved for $\barT$ to yield
\begin{equation}
    \barT=\frac{1}{2} \left(\rho +\sqrt{\rho ^2+\frac{8 \lambda }{\kappa ^2}}\right)\;,
\end{equation}
where the positive square root is favored so that $\barT=\rho$ in the limit where the energy density is large or the coupling $\lambda$ is small. Because of how useful Eq.~(\ref{Quadratic}) is for rewriting the energy density in terms of the stress-energy trace, this equation will henceforth be used throughout our work without referencing it.

Assuming a flat Friedmann-Lemaître-Robinson-Walker metric, i.e. Eq.~({\ref{FLRW Metric}}), the first Friedmann equation can be obtained in terms of $\barT$ by combining Eqs.~(\ref{Einstein EQ}), (\ref{T Variation}), and (\ref{f(R,T)}) for $\epsilon=-1$ to yield
\begin{equation}
    3\left(\frac{\dot{a}}{a}\right)^2=\left(\kappa^2- \frac{5 \lambda }{2\barT^2}\right)\left(\barT-\frac{2\lambda }{\kappa ^2\barT}\right)+\frac{\lambda}{\barT}\;. \label{tt-Friedmann}
\end{equation}
From Eq.~(\ref{ConserveN}), the effective current density
\begin{equation}
    J'^{\mu }=\left(1-\frac{2\lambda }{\kappa ^2\barT^2}\right)J^{\mu }=\frac{\rho }{\barT}\left(n \, u^{\mu }\right)
\end{equation}
must be conserved, i.e. $\nabla _{\mu }J'^{\mu }=0$. Recalling that $\rho\propto n$, this implies that the $\rho^2a^3/\barT$ is constant, which we write as
\begin{equation}
   \frac{\rho ^2}{\barT}a^3=\frac{12N^3}{\kappa^2}\;,
\end{equation}
where the constant was chosen for later comparison and $N$ is an arbitrary normalization factor. 
Rewriting the energy density in terms of the stress-energy trace yields the relations
\begin{align}
    \label{A}a&=\left(\frac{12}{\kappa^2}\right)^{1/3}N\frac{\barT}{\left(\barT^2-\frac{2\lambda}{\kappa^2}\right)^{2/3}} \; , \\
    \label{DotA/A}\frac{\dot{a}}{a}&=-\left(\frac{\dot{\barT}}{\barT}\right)\left(\frac{\kappa^2\barT^2+6\lambda}{3 \kappa^2\barT^2-6\lambda}\right)\;.
\end{align}
It is useful to define an expansion parameter $H_\lambda$, a rescaled time $\tau$ and a rescaled stress-energy trace $x$ by the equations
\begin{subequations}
\begin{align}
    H_\lambda &= \left(\frac{\kappa^2\lambda}{18}\right)^{1/4} \; , \label{Hlambda} \\
    \tau &= H_\lambda t \; , \label{t scaling} \\
    \barT(t)&=x(\tau)\sqrt{\frac{2\lambda }{\kappa ^2}} \; . \label{T Scaling}
\end{align}
\end{subequations}
Substituting Eq.~(\ref{DotA/A}) into (\ref{tt-Friedmann}), and rewriting in terms of the rescaled quantities gives the differential equation
\begin{equation}\label{diff X}
    \frac{dx}{d\tau}=-\frac{3(x^2-1)}{x^2+3}\,\sqrt{\frac{4 x^4-7x^2+5}{2x}}\; ,
\end{equation}
effectively describing the behavior of the energy density.

For the early Universe, the energy density is infinite while for the late Universe, the energy density approaches a constant, i.e. $x\rightarrow\infty$ and $x\rightarrow1+\delta x$ respectively. Solving Eq.~(\ref{diff X}) in these limits yields
\begin{equation}\label{X}
    x(\tau)=
    \begin{cases} 
      \frac{2}{9\tau^{2}} & \tau\rightarrow 0\;, \\
      1+\alpha \exp\left(-\frac32\tau\right)& \tau\rightarrow\infty\;, 
    \end{cases}
\end{equation}
where $\alpha \approx 0.9844$ was determined numerically. This leads to a scale factor which has the asymptotic behavior
\begin{equation}\label{Limit A}
    a(t)=
    \begin{cases} 
      N(3t)^{2/3} & t\rightarrow 0\;, \\
      NH_\lambda^{-2/3} (\sqrt{2}\alpha)^{-2/3}\exp(H_\lambda t)& t\rightarrow\infty\;.
    \end{cases}
\end{equation}
\begin{figure}
\includegraphics[width=\columnwidth]{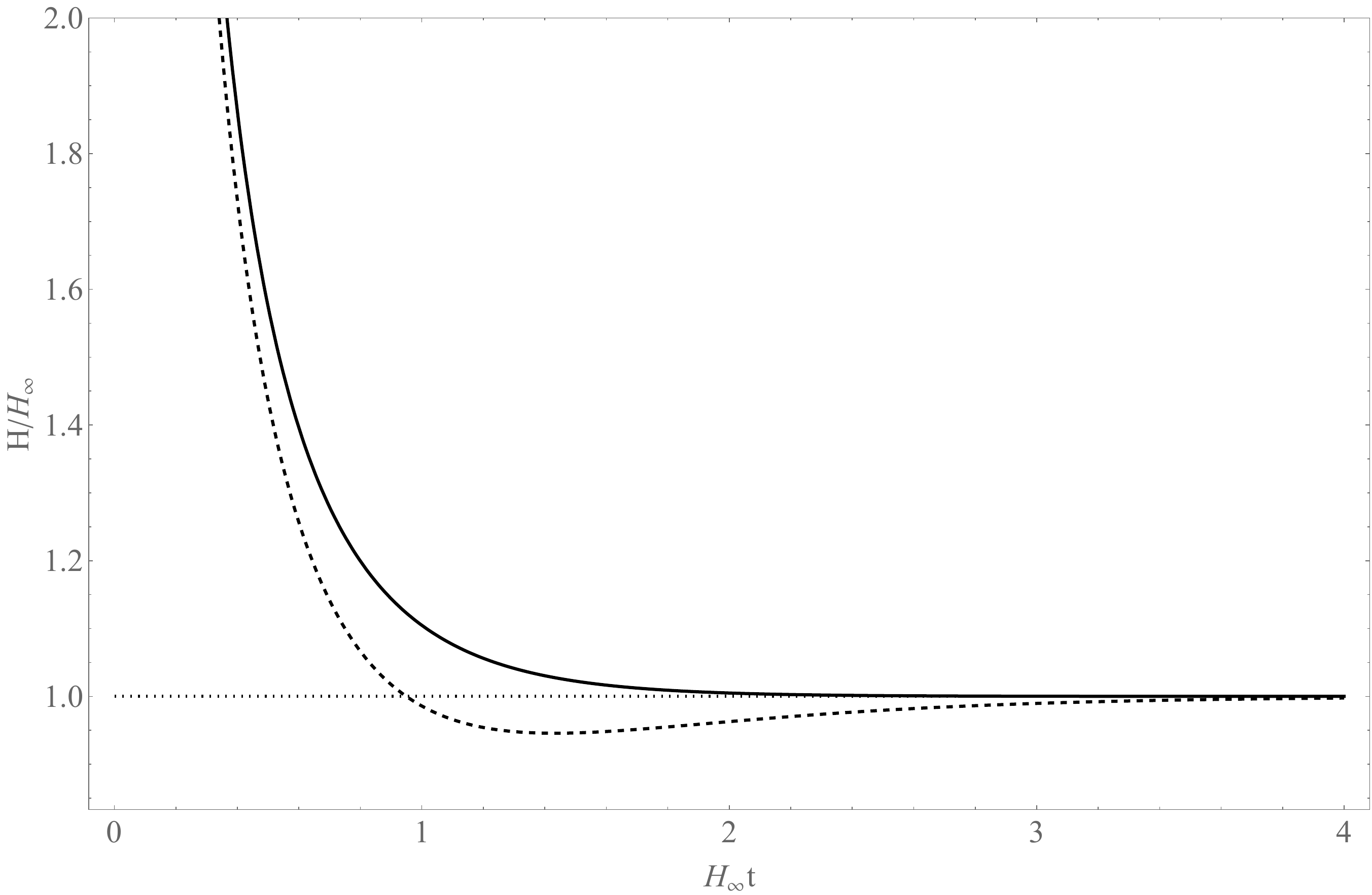}
\caption{\label{fig:H_Constant} The behavior of the Hubble parameter as a function of time in the $\Lambda$CDM (solid) and $f(R,T)$ (dashed) models. The end-time value $H_\infty$ corresponds to $H_\Lambda$ or $H_\lambda$ for $\Lambda$CDM and $f(R,T)$ respectively.}
\end{figure}
We note that this is nearly identical to Eq.~(\ref{Asymtotic lCDM}), the only difference being an overall scale increase in this model that is smaller by a factor of $(\sqrt{2}\alpha)^{-2/3} \approx 0.8021$. The behavior of the Hubble parameter is shown in Fig.~\ref{fig:H_Constant}. However, there is no particular reason to assume the asymptotic Hubble constants $H_\Lambda$ and $H_\lambda$ should match, since the goal of these models is not to make the Universe with particular future behavior, but instead to match the observed redshift-luminosity curves.

\section{Luminosity Distance and Observation}\label{sec:Data}

The relationship between the luminosity distance and redshift can be understood as
\begin{equation}\label{Initial Luminosity Distance}
    d_L=a_0(1+z) \int_t^{t_0}\frac{d t'}{a(t')}\;,
\end{equation}
where $t_0$ is the time value for the present and $a_0=a(t_0)$. In the $\Lambda$CDM model,
\begin{equation}
    H_0 d_L=(1+z)\int_\frac{1}{1+z}^1\frac{d\psi}{\sqrt{\Omega_\Lambda\psi^4+\Omega_m\psi}}\;,
\end{equation}
where $\Omega_\Lambda+\Omega_m=1$. 

Similarly for the considered $f(R,T)$ theory, a relation between the two can be determined starting by combining Eqs. (\ref{DotA/A})-(\ref{diff X}), followed by evaluating at the present time, 
\begin{equation}\label{Time/Tau}
    H_\lambda= H_0 \sqrt{\frac{2x^3}{4x^4-7x ^2+5}}\Bigg|_{\tau=\tau_0}.
\end{equation}
Then by inserting Eqs.~(\ref{A}) in terms of $x(\tau)$ and (\ref{Time/Tau}) into (\ref{Initial Luminosity Distance}) yields
\begin{align}
    \nonumber H_0 d_L=(1+z)&\left[\frac{1}{(x^2-1)^{2/3}}\sqrt{\frac{4x^4-7x ^2+5}{2x}}\,\right]_{\tau=\tau_0}\\
    &\times\int_\tau^{\tau_0}\frac{(x(\tau')^2-1)^{2/3}}{x(\tau')}d\tau'\;,
\end{align}
where
\begin{equation}
    1+z=\left[\frac{x}{(x^2-1)^{2/3}}\right]_{\tau=\tau_0}\frac{(x^2-1)^{2/3}}{x}\;.
\end{equation}
From the luminosity distance $d_L$, the distance modulus $\mu$ can easily be determined using
\begin{equation}
    \mu=5\,\log_{10}\left(\frac{d_L}{10\,\text{pc}}\right)\;.
\end{equation}

\begin{figure}
\includegraphics[width=\columnwidth]{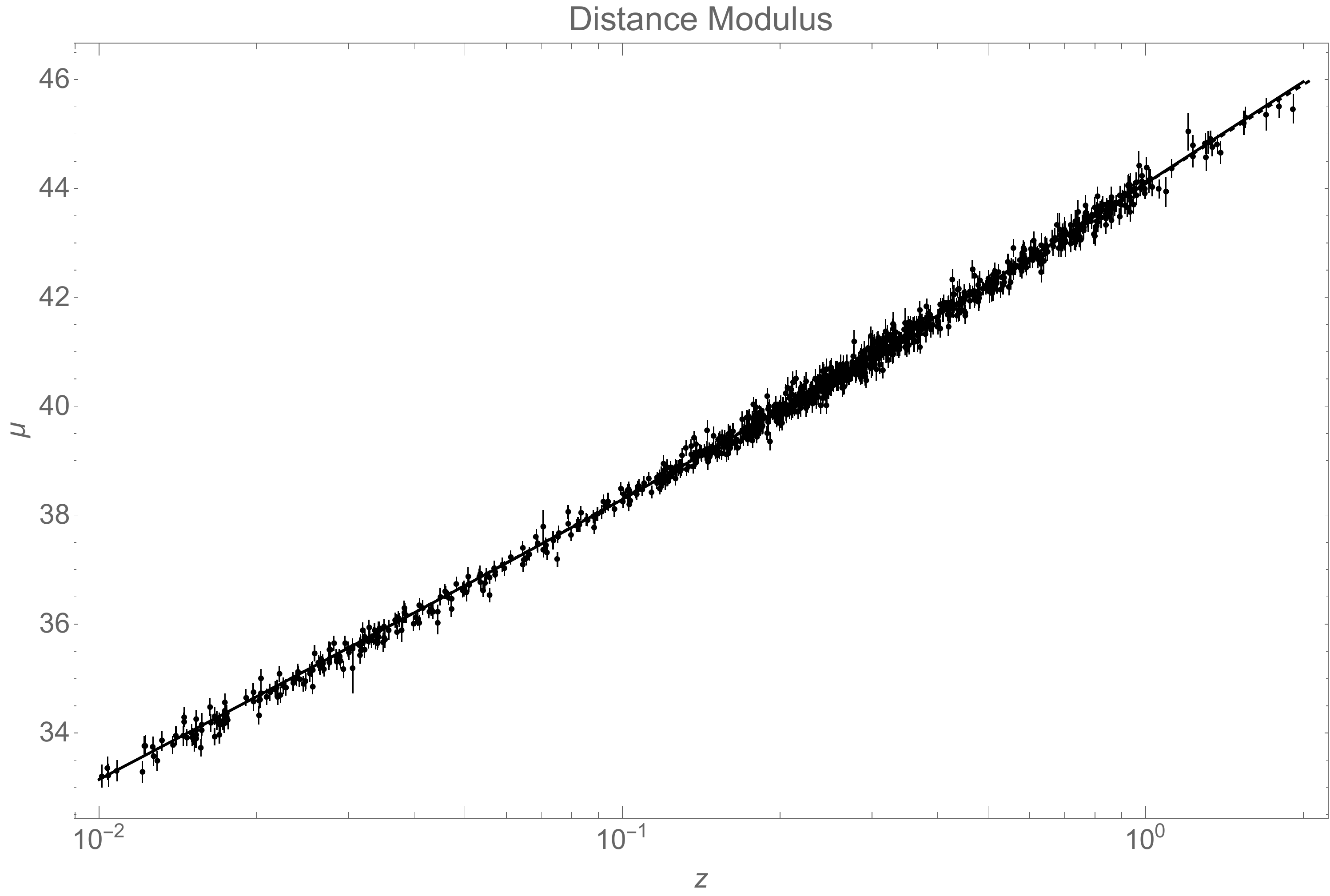}
\caption{\label{fig:DM} The distance modulus $\mu=m-M$ as a function of the redshift factor $z$. The points represent data taken from the Pantheon survey provided by \cite{Scolnic_2018}. The solid line corresponds to the best fit $\Lambda$CDM model. The dashed line represents the best fit of our model at $\tau_0=0.937$.}
\end{figure}

\begin{figure}
\includegraphics[width=\columnwidth]{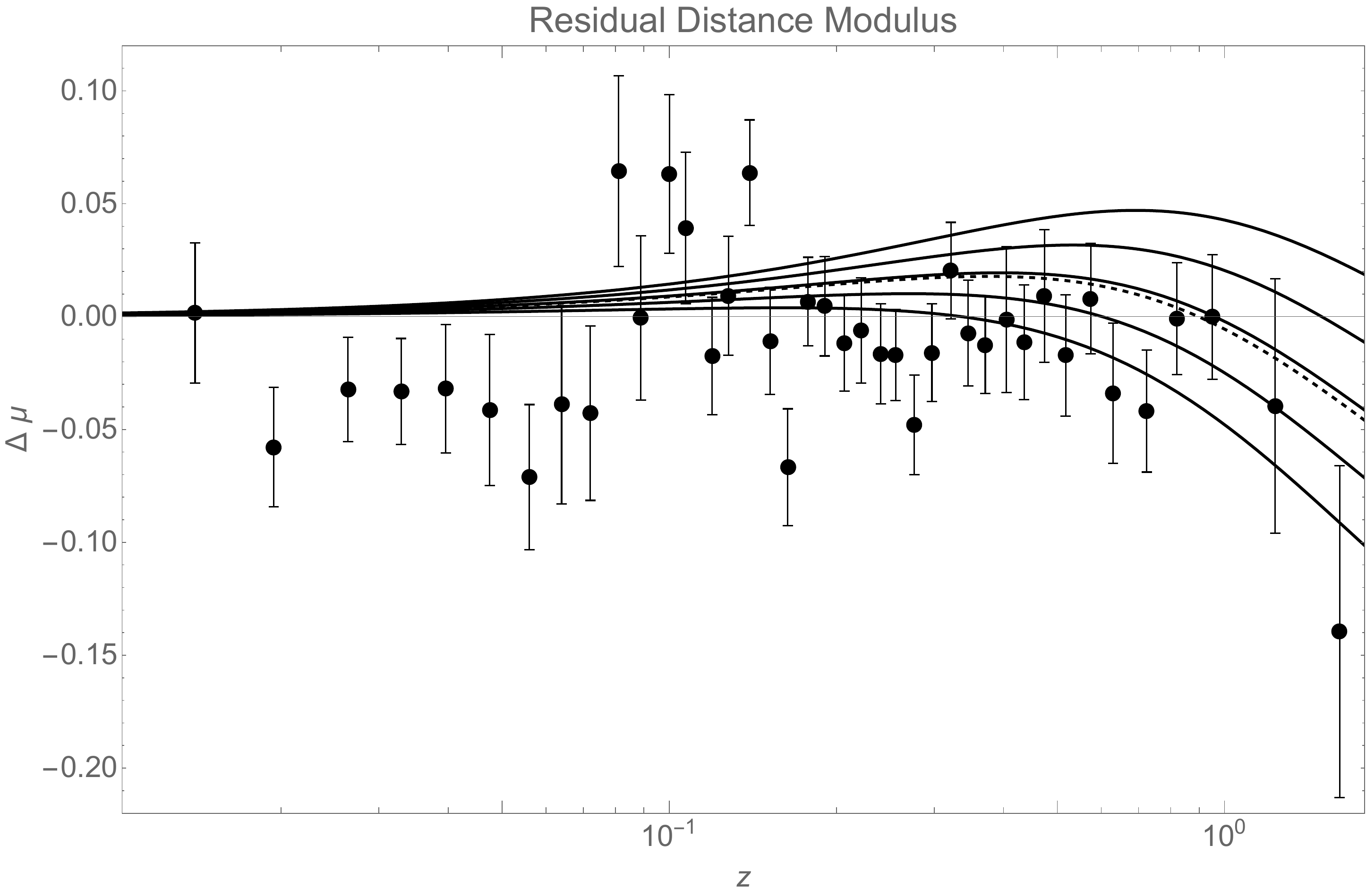}
\caption{\label{fig:Redisual_DM} The difference $\Delta\mu=\mu_f-\mu_\Lambda$ between $f(R,T)$ distance modulus and the best fit standard $\Lambda$CDM modulus. Solid lines correspond to $\tau_0$ values 0.90, 0.92, 0.94, 0.96, and 0.98 from bottom to top. The dashed line represents the best fit of our model at $\tau_0=0.937$. The points represent data taken from the Pantheon binned data provide by \cite{Scolnic_2018}. For fitting purposes, the full Pantheon dataset was used.}
\end{figure}

Figure~\ref{fig:DM} shows the 1048 supernovae data from the Pantheon dataset \cite{Scolnic_2018,Lu_2022,Abbott_2019}, superimposed with the best fit curves for both our model and $\Lambda$CDM. As is clear, the two theories for an appropriate choice of parameters yield virtually indistinguishable curves. To clarify what is going on, we have included Fig.~\ref{fig:Redisual_DM}, which shows the difference between $f(R,T)$ predictions and the best fit $\Lambda$CDM, for different values of $\lambda$. Also included in Fig.~\ref{fig:Redisual_DM} are the ``binned" supernovae data. Again, at least visually, it is not obvious which theory is better. 

To find the best fit values for both our model and $\Lambda$CDM, we minimize $\chi^2$ with respect to $M$, given by
\begin{equation}
    \chi^2=\sum_{i=0}^{n}\left(\frac{\mu(z_i)-m_i+M}{\sigma_i}\right)^2\;,
\end{equation}
where $z_i$, $m_i$, and $\sigma_i$ are the corrected redshift, apparent magnitude, and error in the apparent magnitude of the supernovae data, while $M$ is the absolute magnitude. Then, we scanned through various values of $\tau_0$ and found a minimum for our model at $\tau_0=0.937\pm0.014$. The value of $\lambda$ can be determined by combining Eqs.~(\ref{Hlambda}) and (\ref{Time/Tau}), resulting in
\begin{equation}
    \lambda=\left[\frac{x^6}{(4x^4-7x^2+5)^2}\right]_{\tau=\tau_0}\frac{72 H_0^4}{\kappa^2}\;.
\end{equation}
Subsequently using the best estimate $\tau_0$ yields
\begin{equation}
    \lambda\approx(0.246\pm0.005)\frac{72 H_0^4}{\kappa^2}\;.
\end{equation}
For $H_0=71.5$ km/s/Mpc, this corresponds to $\lambda=5.57(12)\times10^{-76}\;\text{eV}^6$. Our model does not naturally explain the smallness of this parameter; it must simply be treated much like the cosmological constant $\Lambda$ in $\Lambda$CDM.

As in the standard $\Lambda$CDM model, Hubble's constant $H_0$ and the absolute magnitude $M$ for SNe Ia are degenerate parameters \cite{Scolnic_2018}. Recalling that for the $\epsilon=0$ case the $f(R,T)$ model is equivalent to the $\Lambda$CDM model, the second parameter used in both theories is $\lambda$. In both cases there are 1046 degrees of freedom.

Our best fit for the $\Lambda$CDM yields $\chi^2=1035.68$ for the value of $\Omega_\Lambda=0.716$ where $\Omega_\Lambda+\Omega_M=1$, while for our $f(R,T)$ model, multiple $\tau_0$ values were tried in order to find the minimum $\chi^2=1032.64$ at the best fit value of $\tau_0=0.937$. Even though the $\chi^2$ scores suggest our model has a slight advantage in fitting the SNe Ia data, it is not enough to significantly favor our model over $\Lambda$CDM. 

For the best fit values, the two models yield essentially the same values of $M$: at $H_0=71.5$ km/s/Mpc. This results in a best fit value of $ M=-19.32$ for our theory, compared to the conventional $\Lambda$CDM model value of $ M=-19.31$.

\section{Conclusion}

We have studied the expansion of the Universe in $f(R,T)$ gravity with $f(R,T)=R+\lambda T^{-1}$ assuming a flat Universe. Like $\Lambda$CDM, we found an exponential growth transition from the matter-dominated era to present-day cosmology when considering the analysis of \cite{Fisher_2019}. We also compared our predictions for luminosity distance versus redshift with SNe Ia data \cite{Scolnic_2018}. Our results show that this new model is competitive with, or even slightly better than, $\Lambda$CDM.

It is clear from our analysis that supernovae data indicate an accelerating Universe, but current data cannot necessarily distinguish the cause of such acceleration. Since our $f(R,T)$ model fits the data well, as evident in Fig.~\ref{fig:Redisual_DM}, the deviation between standard and nonstandard cosmologies is most evident at high redshift so that more supernovae at high $z$ are needed to differentiate between these models.

In future work, we intend to study alternative models; for example $f(R,T)=R+\lambda T^\epsilon$ for other values of $\epsilon<1$. We also hope to understand how our model affects the fluctuations in the cosmic microwave background radiation and possible modifications in the large scale structure.

\section*{Acknowledgments}

We thank L. Pryor for helpful comments on this paper.

\section*{Data Availability}

The data that support the findings of this article are openly available \cite{Scolnic_2018}.

\medskip

\end{document}